# Quantitative measurement of figure of merit for transverse thermoelectric conversion in Fe/Pt metallic multilayers

Takumi Yamazaki,[1,*,§] Takamasa Hirai,[2,†,§] Takashi Yagi,[3] Yuichiro Yamashita,[3] Ken-ichi Uchida,[1,2] Takeshi Seki,[1,2,‡] and Koki Takanashi[4,5]

[1] *Institute for Materials Research, Tohoku University, Sendai 980-8577, Japan*

[2] *National Institute for Materials Science, Tsukuba 305-0047, Japan*

[3] *National Institute of Advanced Industrial Science and Technology, Tsukuba 305-8563, Japan*

[4] *Advanced Science Research Center, Japan Atomic Energy Agency, Tokai, Ibaraki 319-1195, Japan*

[5] *Advanced Institute for Materials Research, Tohoku University, Sendai, 980-8577, Japan*

[*] takumi.yamazaki.d5@tohoku.ac.jp

[†] HIRAI.Takamasa@nims.go.jp

[‡] takeshi.seki@tohoku.ac.jp

[§] These authors contributed equally to this work.

This study presents a measurement method for determining the figure of merit for transverse thermoelectric conversion ($z_\mathrm{T}T$) in thin film forms. Leveraging the proposed methodology, we comprehensively investigate the transverse thermoelectric coefficient ($S_\mathrm{T}$), in-plane electrical conductivity ($\sigma_{yy}$), and out-of-plane thermal conductivity ($\kappa_{xx}$) in epitaxial and polycrystalline Fe/Pt metallic multilayers. The $\kappa_{xx}$ values of multilayers with a number of stacking repetitions ($N$) of 200 are lower than those of FePt alloy films, indicating that the multilayer structure effectively contributes to the suppression of $\kappa_{xx}$. $z_\mathrm{T}T$ is found to increase with increasing $N$, which remarkably reflects the $N$-dependent enhancement of the $S_\mathrm{T}$ values. Notably, $S_\mathrm{T}$ and $\sigma_{yy}$ are significantly larger in the epitaxial multilayers than those in the polycrystalline counterparts, whereas negligible differences in $\kappa_{xx}$ are observed between the epitaxial and polycrystalline multilayers. This discrepancy in $\sigma_{yy}$ and $\kappa_{xx}$ with respect to crystal growth is due to the different degree of anisotropy in electron transport between epitaxial and polycrystalline multilayers, and epitaxial growth can lead to an enhancement of $z_\mathrm{T}T$ in the multilayers. This study is the first demonstration in the evaluation of $z_\mathrm{T}T$ in thin film forms, and our proposed measurement technique reveals the transverse thermoelectric properties inherent to multilayers.

## I. INTRODUCTION

Transverse thermoelectric conversion refers to the generation of electric field by perpendicularly applying a temperature gradient [1,2]. In contrast to the longitudinal thermoelectric conversion based on the Seebeck effect (SE), which needs three-dimensional serial junctions to achieve high device performance, the transverse thermoelectric conversion has an advantage of a simple device structure without junctions. This simplicity in structure contributes to efficient thermal energy harvesting by minimizing electrical and thermal contact resistance. The anomalous Nernst effect (ANE) [3–7] is a representative transverse thermoelectric phenomenon observed in a magnetic material. Unlike the ordinary Nernst effect [8], which requires a substantial magnetic field to obtain large transverse thermopower, the ANE offers the distinct advantage of generating transverse thermopower under a small magnetic field. It is noteworthy that for magnetic materials with spontaneous magnetization, including permanent magnets, transverse thermopower can be generated in the absence of an external magnetic field [9–11]. The ANE-driven electric field $\mathbf{E}_{\mathrm{ANE}}$ is expressed as

$$\mathbf{E}_{\mathrm{ANE}} = S_{\mathrm{ANE}}\left(\frac{\mathbf{M}}{|\mathbf{M}|} \times \nabla T\right), \tag{1}$$

where $S_{\mathrm{ANE}}$ is the ANE coefficient, $\nabla T$ is the temperature gradient applied to the magnetic material, and $\mathbf{M}$ is the magnetization [1]. $S_{\mathrm{ANE}}$ can be expressed as

$$S_{\mathrm{ANE}} = \rho_{xx}\alpha_{xy} - \rho_{yx}\alpha_{xx}, \tag{2}$$

where $\rho_{xx}$, $\rho_{yx}$, $\alpha_{xx}$, and $\alpha_{xy}$ refer to the longitudinal resistivity, anomalous Hall resistivity, longitudinal thermoelectric conductivity, and transverse thermoelectric conductivity, respectively [6]. The first term on the right-hand side of Eq. (2) refers to the direct conversion from $\nabla T$ to $\mathbf{E}_{\mathrm{ANE}}$ due to $\alpha_{xy}$, which is attributed to the Berry curvature near the Fermi energy in electronic band structures and/or spin-dependent disorder scattering. On the other hand, the second term on the right-hand side of Eq. (2) originates from the anomalous Hall effect of the longitudinal carrier flow induced by the SE. In addition to the characteristic symmetry of transverse thermoelectric conversion, it has a potential of controlling the sign of $\mathbf{E}_{\mathrm{ANE}}$ by changing the direction of $\mathbf{M}$ as shown in Eq. (1), enabling us to construct a thermopile structure within a single magnetic material [12]. Although the ANE possesses distinct advantages over the SE as mentioned above, the improvement of the performance of thermoelectric conversion is a crucial issue to realize practical thermoelectric applications.

The performance of the ANE-induced transverse thermoelectric conversion is evaluated using the dimensionless figure of merit $z_{\mathrm{ANE}}T$ (= $S_{\mathrm{ANE}}^2\sigma_{yy}T/\kappa_{xx}$), where $\sigma_{yy}$, $\kappa_{xx}$, and $T$ denote the electrical conductivity along $\mathbf{E}_{\mathrm{ANE}}$, thermal conductivity along $\nabla T$, and absolute temperature, respectively. For an efficient transverse thermoelectric conversion, it is essential to obtain high $S_{\mathrm{ANE}}$, high $\sigma_{yy}$, and low $\kappa_{xx}$. Currently, the mainstream of materials research for the ANE is developing materials with high $S_{\mathrm{ANE}}$. So far, some topological ferromagnets and non-collinear antiferromagnets have been found to exhibit high $S_{\mathrm{ANE}}$ values owing to the large Berry curvature near the Fermi level [13–21]. The record-high $S_{\mathrm{ANE}}$ value above room temperature is 6-8 μV K$^{-1}$ for Co$_2$MnGa [15,19], and that in low temperature region is 23 μV K$^{-1}$ at 40 K for UCo$_{0.8}$Ru$_{0.2}$Al [20]. These materials also exhibit a large $z_{\mathrm{ANE}}T$ of about 0.5×10$^{-3}$ at 300 K for Co$_2$MnGa and about 1.6×10$^{-3}$ at 40 K for UCo$_{0.8}$Ru$_{0.2}$Al, which were summarized in Ref. [21]. Permanent magnet SmCo$_5$, which has a unique advantage of operating under zero-magnetic field due to its large remanent magnetization and coercivity, achieves large $S_{ANE}$ of 5.6 μV K$^{-1}$ and $z_{ANE}T$ of $1.1 \times 10^{-3}$ at 572 K [9,22]. Bulk single-crystalline YbMnBi$_2$ shows a large $S_T$ of about 3 and 6 μV K$^{-1}$ at around 170 K, depending on its crystallographic orientation [21]. It also has a large $z_{ANE}T$ value of about $3 \times 10^{-3}$ at around 170 K when a heat flux is applied to a specific crystallographic orientation, which is attributed not only to the substantial $S_{ANE}$ but to the layered structure that induces anisotropy in the transport properties [23]. In other words, the anisotropy of the electrical and thermal transport properties plays an important role in enhancing the performance of transverse thermoelectric conversion.

Another promising candidate to achieve high transverse thermopower is multilayers. Several studies have reported that the transverse thermopower can be enhanced by forming the multilayer structures with a number of interfaces [24–29]. It should be noted that in the in-plane magnetized configuration of multilayers, in addition to the ANE, the spin Seebeck effect (SSE) [30–33] and/or the spin-dependent Seebeck effect (SdSE) may affect the transverse thermopower [34]. In the case of multilayers with in-plane magnetized configuration, thus, we rewrite the transverse thermoelectric coefficient to be $S_T$. The SSE and SdSE refer to the heat-to-spin current conversion phenomena in a magnet/heavy metal junction driven by magnon spin current and conduction electron spin current, respectively. When combined with the inverse spin Hall effect in the heavy metal layer [35], these effects exhibit the same thermoelectric conversion symmetry as the ANE in the in-plane magnetized configuration [36]. In the case of the in-plane magnetized configuration, i.e., the direction of electrical conduction ($y$) is parallel to the in-plane of multilayer, and the direction of heat conduction ($x$) is perpendicular to the superlattice surface as illustrated in Fig. 1, the inherent structural anisotropy of multilayers is expected to induce the anisotropy of transport properties. That is, one may anticipate that $\kappa_{xx}$ is reduced remarkably by forming the multilayer structure while the reduction of $\sigma_{yy}$ is not so significant, resulting in an enhancement of the dimensionless figure of merit for the transverse thermoelectric conversion $z_T T$ ($= S_T^2 \sigma_{yy} T / \kappa_{xx}$) in multilayers. However, the influence of formation of multilayer on $\kappa_{xx}$ has not experimentally been examined yet using the systematic samples due to the difficulty in the thermal conductivity measurements for thin films. Moreover, the quantitative evaluation for $S_T$ is not established for thin film samples with the in-plane magnetized configuration, which leads to the lack of report on the evaluation of $z_T T$ for multilayers. The following paragraph explains this technical issue in detail.

Thin films, including multilayers, offer advantages for practical thermoelectric application, such as in micromachines and wearable devices, whereas the evaluation of their thermoelectric figure of merit is quite challenging. In the case of SE, thin films have garnered increasing attention because it has been demonstrated that the multilayer structure and high-density two-dimensional electron gas can improve the thermoelectric conversion performance of thin films in comparison to bulk materials [37]. It is essential for the SE to align the directions of charge and heat currents to prevent an overestimation of the figure of merit. When a heat current is applied to the in-plane direction, the in-plane thermal conductivity is hard to be quantified because of heat leakage to the substrate. The quantitative evaluation of in-plane thermal conductivity has only been established in restricted structures, such as suspended or supported films [38]. On the other hand, when a heat current is directed perpendicular to the plane, evaluating the out-of-plane Seebeck coefficient poses challenges. Several recent reports have proposed a measurement method for the out-of-plane Seebeck coefficient, but this method does not account for the influence of contact/interfacial thermal resistances when measuring the temperature difference between the top and bottom surfaces of the thin-film sample [39,40]. In contrast to the SE, since the ANE allows the orthogonal configuration between the heat and charge currents, it is sufficient to evaluate $S_T$ and $\sigma_{yy}$ in the in-plane direction and $\kappa_{xx}$ in the out-of-plane direction. Various measurement methods, including optical and electrical techniques, have been well-established to quantify $\kappa_{xx}$ of thin films in the out-of-plane direction [41–45], which is advantage of the ANE case. Nevertheless, no reports have been published to date on the $z_T T$ evaluation for thin film forms.

In this study, we propose and demonstrate the method accurately evaluating $z_T T$ in thin film forms. The values of $S_T$, in-plane $\sigma_{yy}$, and out-of-plane $\kappa_{xx}$ are investigated for the Fe/Pt metallic multilayers with the in-plane magnetized configuration. Samples were grown on two different substrates, which resulted in two different growth types: epitaxial and polycrystalline multilayers. These multilayers allow us to examine the effect of epitaxial growth on transport properties. To obtain out-of-plane $\kappa_{xx}$, the time-domain thermoreflectance (TDTR) method [42,43] was employed, which is a well-established technique for characterizing micro/nano-scale thermal transport, and is known for its ability to evaluate the effective out-of-plane $\kappa_{xx}$ for metallic multilayers [46–48]. The value of $S_T$ for the in-plane magnetized configuration was measured by combining the heat flux method [49,50] and the $\kappa_{xx}$ value from the TDTR measurement. The inherent structural anisotropy of multilayers leads to the difference in the electron transport properties between the in-plane and out-of-plane directions, which is evaluated as the ratio of electrical and thermal transport properties $\sigma_{yy}/\kappa_{xx}$. Finally, we systematically investigate the $z_T T$ value as a function of number of stacking repetitions $N$ for the epitaxial and polycrystalline multilayers.

## II. EXPERIMENTAL PROCEDURE

Fe/Pt metallic multilayers were deposited on single-crystalline $SrTiO_3$ (STO)(001) and amorphous quartz glass (hereinafter quartz for simplicity) substrates using an ultrahigh-vacuum-compatible magnetron sputtering system with a base pressure of $\sim 10^{-7}$ Pa. The layer stackings from bottom to top are Fe(2 nm)/[Pt($t_{Pt}$)/Fe($t_{Fe}$)]$_N$ with $N$ = 50, 100, and 200, where $t_{Pt}$ and $t_{Fe}$ are the unit thicknesses of Pt and Fe layers in the multilayer structure, respectively. Figure 2(a) displays the illustration of layer stacking of the present multilayers. Prior to the sputter-deposition, the STO substrate was heated up to 450°C for surface cleaning while that substrate heating was not carried out for the quartz substrates. Then, all the Fe and Pt layers were deposited at room temperature. For the samples of TDTR measurement, the Al layer was deposited on the Fe/Pt multilayers at room temperature. On the other hand, the samples for measuring $S_T$ and $\sigma_{yy}$ do not have an Al layer in order to prevent the electrical shunting problem. In addition to the Fe/Pt multilayers, the samples with the FePt alloy layers were prepared by co-deposition of Fe and Pt at room temperature as reference samples, where the FePt is a disordered FePt because of the room temperature deposition without any thermal treatment. Accurate determination of the Al thickness $t_{Al}$, the total thickness of Fe/Pt multilayer $t_{ML}$, and the thickness of Fe-Pt alloy film $t_{alloy}$ is essential because the TDTR measurements are exceedingly sensitive to these parameters. In order to determine these thicknesses, we performed cross-sectional scanning transmission electron microscopy (STEM) for the samples on STO. The structures of the Fe/Pt multilayers and FePt alloy films were characterized using X-ray diffraction (XRD).

The values of $\kappa_{xx}$, $S_T$, and $\sigma_{yy}$ were measured by employing the TDTR method, heat flux method, and conventional four-terminal method, respectively. Details of each method are as follows. TDTR is an optical pump-probe method that measures thermal transport properties by detecting the transient surface temperature modulation via thermoreflectance. The TDTR setup used in this study is based on front-heating and front-detection configuration, where the repetition rate, pulse duration, central wavelength, and $1/e^2$ spot diameter of pump (probe) laser were 20 MHz (20 MHz), 0.5 ps (0.5 ps), 1550 nm (775 nm), and ~90 μm (~30 μm), respectively. The laser powers were determined so that the steady-state temperature rising at the sample surface was < 2 K [42]. Since the change in the surface reflectivity due to thermoreflectance is small (at most $10^{-2}$%), a lock-in detection scheme was conducted as following: the pump laser modulated at a frequency of 200 kHz was irradiated on the sample surface, the probe laser was focused on the same spot with a specific time delay which was electrically controlled by a function generator [10,51], and then the lock-in signals synchronized with the modulation frequency due to surface temperature response were characterized in the time domain by measuring the reflected probe laser using a balanced photoreceiver connected to a lock-in amplifier. The magnitudes of $\kappa_{xx}$ were determined by a numerical simulation based on the one-dimensional heat diffusion model [52]. This approach is suitable because the spot diameter of pump laser significantly exceeds the heat diffusion length (approximately 2.5 μm) during the modulation period in the STO substrate, and consequently, heat can be regarded as flowing exclusively in the out-of-plane direction, thereby eliminating the need to consider in-plane heat conduction. In the heat diffusion model, the Fe/Pt multilayer was regarded as a single homogeneous layer, and thus the obtained $\kappa_{xx}$ is the effective value for the multilayer, where the contributions of the Fe and Pt layers and Fe/Pt interfaces are included. For TDTR analysis, we used the lock-in phase value $\varphi_{TR}$, corresponding to the ratio of in-phase and out-of-phase components of lock-in signal, according to the general analyses [42]. Note that the $\varphi_{TR}$ data at the delay time of > 0.1 ns were used for analysis, where the electron-phonon coupling effect in the Al transducer layer dies out. $S_T$ for the in-plane magnetization configuration was evaluated using the heat flux method. It is known as a method with high reproducibility compared with the temperature difference method [50], which was demonstrated in the measurement of the SSE. This method gives the value of transverse electric field $E_T$ normalized by the applied heat flux $-j_q$. Here, it is noted that $E_T$ is derived from the voltage measurement at an open circuit condition and the sign of $E_T$ is opposite to that of $E_{ANE}$ [1]. $S_T$ can be quantitatively estimated using the relation of $S_T = \kappa_{xx}E_T/(-j_q)$ derived from Eq. (1) and Fourier's law. The details of the heat flux method are shown in the next section. An important point is that the $\kappa_{xx}$ values can be obtained from the TDTR measurements as mentioned above. For the $\sigma_{yy}$, measurement, the standard four-terminal method was performed for the microfabricated devices. The samples were patterned into Hall bars through photolithography and Ar ion milling processes. The Cr/Au electrodes were fabricated using ion-beam sputtering and lift-off process. These measurements for transport properties were performed at room temperature and atmospheric pressure. In the heat flux method, the effects of heat loss due to radiation and convection were neglected.

## III. RESULTS AND DISCUSSION

Clear multilayer structures of the samples on STO are observed from the cross-sectional STEM images. Figure 2(b) shows bright field (BF)-STEM images for the Fe/Pt multilayers on STO with $N$ = 50, 100, and 200. We experimentally obtain $t_{Al}$ and $t_{ML}$ ($t_{alloy}$) for Fe/Pt multilayers (FePt alloy film) from BF-STEM images, which are summarized in Table 1. The high angle annular dark field (HAADF)-STEM images with higher resolution [see Fig. 2(c)] also confirmed that all of the Fe/Pt multilayer samples successfully form the layer stacking with clear interfaces.

Figure 3 shows out-of-plane XRD patterns for (a) the Fe/Pt multilayers on STO and (b) quartz substrates. The XRD patterns for the FePt alloy films are also shown for comparison. All the Fe/Pt multilayers exhibit multiple peaks, and these diffraction patterns are reproduced by the calculation based on the Laue function with the assumption of the step model for the superstructure (see Supplemental Material [53]). From these XRD patterns and the reflection high-energy electron diffraction patterns (see Supplemental Material [53]), we found that the Fe/Pt multilayers and FePt layer on STO are epitaxially grown with the (100) preferential crystal orientation whereas those on quartz show the polycrystalline growth.

Figure 4(a)[4(b)] shows $\varphi_{TR}$ for the Fe/Pt multilayers and the FePt alloy film on STO (quartz) substates as a function of the delay time between pump and probe lasers. The experimental data were well fitted by the one-dimensional heat diffusion model (black solid lines). The detailed information of TDTR analysis is described in Supplemental Material [53]. The estimated $\kappa_{xx}$ values are summarized in Fig. 4(c). With increasing $N$ of the multilayer film on both substrates, the $\kappa_{xx}$ value decreases. The difference in substrate, i.e., the difference in crystalline orientation and/or grain size does not noticeably affect the magnitude of $\kappa_{xx}$ when comparing samples with the same $N$. Importantly, $\kappa_{xx}$ of the Fe/Pt multilayers with $N = 200$ is apparently lower than that of FePt alloy films [shown as horizontal lines in Fig. 4(c)], proving that arrangement of multilayer structure effectively suppresses $\kappa_{xx}$ for ANE-based thermoelectric devices.

Figure 5(a) illustrates the setup to measure $S_T$. A chip heater with a resistance of 1 k$\Omega$ and a size of 3.02×1.54×0.41 mm, the sample with a size of about 5×2×0.5 mm, and a sapphire slab with contact pads were stacked on a heat sink. The chip heater and sample were fixed with varnish, whereas the sapphire slab was fixed with a thermally conductive silicone adhesive sheet. The Al-1% Si bonding wires with a diameter of 25 μm (Cu wires with a diameter of 50 μm) were used for electrical connection between the chip heater (the sample) and contact pads to minimize the thermal conduction loss. The Cu wires were rigidly connected to the sample via indium soldering and the distance between the indium contacts was adjusted to the $y$-directional length of the chip heater $l$ (= 3.02 mm). The measurement procedure is as follows; a heat flux $\mathbf{j}_q$ was produced along the out-of-plane $x$ direction through the sample by applying the power to the heater and the steady state of the system was achieved, then a magnetic field $\mathbf{H}$ was applied to the in-plane $z$ direction, and a voltage $V_y$ was measured along the in-plane $y$ direction using a nanovoltmeter. A temperature rise $\Delta T$ of the heater surface from room temperature as a function of $-j_q$ was confirmed for the samples on STO and quartz substrates [Fig. 5(b)]. Here, the heater surface was covered with a black ink with emissivity of > 0.94, and $\Delta T$ was measured using an infrared camera. $\Delta T$ shows a linear response to $-j_q$ and the magnitude of $\Delta T$ reflects the difference in thermal conductivity of substrates. Figure 5(d) displays the $H$ dependence of $V_y$ for the Fe/Pt multilayer on STO with $N = 200$ for $-j_q = 48.4$ kW m$^{-2}$. The background contribution of induced electromotive force due to electromagnetic induction is subtracted in advance using a dataset for $-j_q = 0$. It clearly exhibits the $H$-odd function in response to its magnetization reversal [Fig. 5(c)]. The value of $E_T$ is derived from the relation $E_T = (V_y^+ - V_y^-)/2l$, where $V_y^{+(-)}$ is extrapolated from positive (negative) high magnetic field in the region of $|H|$ = 6-10 kOe to zero magnetic field. Note that $|H|$ = 6 kOe is much larger than the saturation magnetic field for all the samples. Figures 5(e) and 5(f) show the $E_T$ values as a function of $-j_q$ for the samples on STO and quartz, respectively. The $E_T/(-j_q)$ values are evaluated from the slope of the linear fit, and $S_T$ was obtained using the experimentally obtained $E_T/(-j_q)$ and $\kappa_{xx}$.

Figure 6(a) shows the $N$ dependence of $S_T$ for the Fe/Pt multilayers on STO and quartz. The $S_T$ values for FePt alloy films are shown as horizontal lines for comparison. $S_T$ increases monotonically with increasing $N$, and tends to be larger than that for the samples on quartz. The FePt alloy film on quartz exhibits the maximum $S_T$ of (1.7±0.1) μV K$^{-1}$, which is of the same order as the ANE coefficient of partially $L1_0$-ordered FePt alloy film for the out-of-plane magnetized configuration [54,55]. The enhancement of $S_T$ with increasing $N$ is consistent with previous studies on the ANE for multilayers [24,25]. Possible scenarios for explaining the enhancement of the ANE in this system are (i) the interdiffusion and alloying of Fe and Pt at the interfaces, (ii) the contribution of proximity-induced magnetic moments in Pt [24], and (iii) the enhancement of spin-orbit interaction at the interfaces. The well-defined structures of Fe/Pt multilayers shown in Fig. 2(c) suggest that there are few regions of intermixing between the Fe and Pt layers. Even if the $S_T$ in ultra-thin intermixed/alloyed layer is as large as that in the FePt alloy layer, the substantial $S_T$ enhancement cannot be attributed solely to the intermixing and alloying of Fe and Pt at the interfaces since the presence of shunting at remnant Pt layers would reduce the thermopower. Hence, scenario (i) does not account for the enhancement of $S_T$. In addition, it has been demonstrated that the ANE for the Fe/nonmagnetic metal multilayers is enhanced regardless of the presence or absence of proximity ferromagnetism [24], which rules out scenario (ii) as the main contributing factor for the increase in $S_T$ with $N$ for this system. At this moment, we believe that scenario (iii) plays a crucial role in achieving this enhancement. One may think that the SSE and SdSE could potentially contribute to enhancing $S_T$ in the Fe/Pt multilayer because both of them occur in a magnet/heavy metal junction in in-plane magnetized configuration. It should be emphasized that the SSE and SdSE cannot be experimentally separated from the ANE in the present measurement setup because they have the same thermoelectric symmetry as the ANE in the case of in-plane magnetized configuration. Moreover, the SSE can be enhanced by multilayer structure [56], whereas the SSE and SdSE are substantially affected by factors such as the thickness of Pt and Fe and the shunting effect to Fe layer. Therefore, it is difficult to quantitatively evaluate the contributions of the SSE and SdSE to the enhancement of $S_T$. Nonetheless, it is quite possible that $S_T$ is enhanced by the contribution of the ANE alone. This is supported by the report that $S_T$ increases with increasing $N$ in the Fe/Pt multilayers even for the out-of-plane magnetized configuration, in which the SSE and SdSE do not appear [24].

An interesting point is that the Fe/Pt multilayer with $N = 200$ on STO shows the value of $S_T$ nearly comparable to that of FePt alloy on STO. Each layer thickness for $N = 200$ is $t_{Fe} \sim 0.5$ nm and $t_{Pt} \sim 0.7$ nm, which is confirmed from Fig. 2(c). These thicknesses are still sufficiently larger than the unit cell size of FePt ($\sim 0.38$ nm), and the local formation of FePt, for example at the interface, is not responsible for the value of $S_T$. These facts mean that the origin of $S_T$ in Fe/Pt multilayers is a hybrid of several contributions, including the bulk ANE in Fe and Pt layers, interfacial ANE, SSE, and SdSE. These multiple origins are in contrast to the $S_T$ in the FePt alloy film, which originates solely from the bulk ANE in the FePt alloy layer. It is particularly interesting that despite the fundamentally different origins of the transverse thermoelectric conversion, the $S_T$ values for the Fe/Pt multilayer with $N = 200$ and FePt alloy film are comparable. In order to fully understand the connection between them, further investigation, including quantitative separation of the individual contributions within the Fe/Pt multilayer, is needed.

Subsequently, we investigate the $N$ dependence of $\sigma_{yy}$ for the Fe/Pt multilayers. Figure 6(b) plots the $N$ dependence of $\sigma_{yy}$ for the Fe/Pt multilayers on STO and quartz. $\sigma_{yy}$ is monotonically decreased with increasing $N$, and the minimum value is obtained at $N = 200$ for both substrates. Interestingly, while the $\sigma_{yy}$ value for FePt alloy film shows minimum in the case of the sample on quartz, the $\sigma_{yy}$ value for the FePt alloy film on STO is higher than that for the Fe/Pt multilayers on STO with $N = 100$ and 200. One possible explanation for this discrepancy is the reduced scattering of electrons due to relaxed lattice distortion in the epitaxial FePt alloy.

While the dependence of $\sigma_{yy}$ and $\kappa_{xx}$ on $N$ exhibits a similar decreasing behavior with increasing $N$, a discrepancy appears when the results are compared between the samples on STO and quartz substrates. To provide a detailed discussion of this discrepancy, the ratio of $\sigma_{yy}$ to $\kappa_{xx}$, i.e., $\sigma_{yy}/\kappa_{xx}$ as a function of $N$ was plotted in Fig. 6(c). In both multilayers on STO and quartz, the $\sigma_{yy}/\kappa_{xx}$ value reaches its maximum at $N = 50$ and subsequently diminishes to a nearly constant value at $N = 100$ and 200. This behavior is attributed to the influence of electron scattering at interfaces in the multilayer. Considering the anisotropic structure of multilayer, one may anticipate that the electrons moving in the out-of-plane direction are remarkably scattered at the interface, resulting in the anisotropy in electron transport between the in-plane and out-of-plane directions. As the thickness of each layer decreases, however, interfacial electron scattering affects electron transport in the in-plane direction as well as that along the out-of-plane direction, leading to less anisotropic behavior between the transport properties along the in-plane and the out-of-plane directions. The multilayers on STO exhibits larger $\sigma_{yy}/\kappa_{xx}$ values than those on quartz, with the attenuation behavior being more pronounced in the former. This behavior is ascribed to the high crystalline orientation and/or large grain size of the epitaxial multilayers on STO in comparison to the polycrystalline multilayers on quartz. This allows interfacial electron scattering to exert a more prominent influence in the epitaxial multilayers, leading to boosting the anisotropy in the transport properties. Considering the expression for $z_T T$, one can see that greater anisotropy between $\sigma_{yy}$ and $\kappa_{xx}$, i.e., the higher $\sigma_{yy}/\kappa_{xx}$ value, leads to a higher $z_T T$. Therefore, to improve the transverse thermoelectric conversion performance by utilizing the structural anisotropy of the multilayer, it is essential to grow the multilayer epitaxially and optimize the thickness of each layer to maintain in-plane $\sigma_{yy}$.

Finally, the $z_T T$ values for the Fe/Pt multilayers are evaluated using the obtained $S_T$, $\sigma_{yy}$, and $\kappa_{xx}$ values. Figure 6(d) shows the $N$ dependence of $z_T T$ of the Fe/Pt multilayers at $T = 296$ K. The $z_T T$ value is increased with increasing $N$, and displays a consistently large value for the multilayers on STO compared to those on quartz. While the $\sigma_{yy}/\kappa_{xx}$ value shows its maximum in the multilayer with $N = 50$ as shown in Fig. 6(c), the substantial contribution from $S_T$ achieves a maximum $z_T T$ value at $N = 200$. Considering that the epitaxial growth of multilayer hardly affects $\kappa_{xx}$ and strongly affects the enhancement of $S_T$ and $\sigma_{yy}$, the epitaxial growth can be an effective strategy to increase the $z_T T$ of multilayers. It is worth mentioning that this study represents the pioneering effort in demonstrating the evaluation of $z_TT$ in thin film form.

## IV. CONCLUSION

We demonstrated a measurement method for determining $z_TT$ value of thin films in the in-plane magnetized configuration. Utilizing our proposed methodology, we conducted a comprehensive investigation for the $N$ dependence of $S_T$, $\sigma_{yy}$, and $\kappa_{xx}$ in the epitaxial and polycrystalline Fe/Pt metallic multilayers. The $\kappa_{xx}$ value in the out-of-plane direction was measured using the TDTR method. $S_T$ for the in-plane magnetization configuration and $\sigma_{yy}$ were evaluated using the heat flux method and conventional four-terminal method, respectively. The combination of the heat flux method and the TDTR method enabled us to evaluate the $S_T$ value for the in-plane magnetized configuration with high reproducibility, which could not be achieved using the conventional temperature difference method. The $N$ dependence of $z_TT$ increased with increasing $N$, which remarkably reflected the $N$ dependent behavior of the $S_T$ values. $S_T$ and $\sigma_{yy}$ were notably larger in the epitaxial multilayers than those for the polycrystalline multilayers, whereas negligible differences in $\kappa_{xx}$ were observed between the epitaxial and polycrystalline multilayers. This difference between the epitaxial and polycrystalline multilayers was further confirmed in the $\sigma_{yy}/\kappa_{xx}$ values, which reflected the degree of electron transport anisotropy. Taking into consideration the consistently larger $S_T$ and $\sigma_{yy}/\kappa_{xx}$ values observed in epitaxial films, leading to enhanced $z_TT$, it is suggested that epitaxial growth is a more efficient route than polycrystalline growth for ANE-based transverse thermoelectric conversion in multilayers. This study represents a pioneering effort in the evaluation of $z_TT$ in thin film form. Our proposed measurement technique unveils insights into the transverse thermoelectric properties inherent to multilayers and provides guidelines for designing multilayers that exhibit high transverse thermoelectric conversion performance: the balance of high anisotropy in transport properties ($\sigma_{yy}/\kappa_{xx}$) and large $S_T$. We anticipate that this approach will greatly facilitate materials research in the field of thermoelectricity and spin caloritronics.

## ACKNOWLEDGEMENTS

The authors thank Y. Sakuraba for his help of supporting ANE measurements, M. Nagasako for his help of scanning transmission electron microscopy, and T. Sasaki for her help with the film deposition by ion beam sputtering. The device fabrication was partly carried out at the Cooperative Research and Development Center for Advanced Materials, Institute for Materials Research, Tohoku University. This work was supported by Grant-in-Aid for Scientific Research (S) (Grant No. JP18H05246 and JP22H04965) and Grant-in-Aid for Research Activity Start-up (Grant No. JP22K20495) from JSPS KAKENHI, Japan; CREST "Creation of Innovative Core Technologies for Nano-enabled Thermal Management" (Grant No. JPMJCR17I1) and ERATO "Magnetic Thermal Management Materials" (Grant No. JPMJER2201) from JST, Japan. T. Y. is supported by JSPS through a Research Fellowship for Young Scientists (Grant No. JP22KJ0210).

[1] K. Uchida, W. Zhou, and Y. Sakuraba, Transverse thermoelectric generation using magnetic materials, Appl. Phys. Lett. **118**, 140504 (2021).

[2] K. Uchida and J. P. Heremans, Thermoelectrics: From longitudinal to transverse, Joule **6**, 2240 (2022).

[3] T. Miyasato, N. Abe, T. Fujii, A. Asamitsu, S. Onoda, Y. Onose, N. Nagaosa, and Y. Tokura, Crossover behavior of the anomalous Hall Effect and anomalous Nernst effect in itinerant ferromagnets, Phys. Rev. Lett. **99**, 086602 (2007).

[4] Y. Pu, D. Chiba, F. Matsukura, H. Ohno, and J. Shi, Mott Relation for Anomalous Hall and Nernst Effects in $Ga_{1-x}Mn_xAs$ Ferromagnetic Semiconductors, Phys. Rev. Lett. **101**, 117208 (2008).

[5] Y. Sakuraba, K. Hasegawa, M. Mizuguchi, T. Kubota, S. Mizukami, T. Miyazaki, and K. Takanashi, Anomalous Nernst Effect in $L1_0$-FePt/MnGa Thermopiles for New Thermoelectric Applications, Appl. Phys. Express **6**, 033003 (2013).

[6] Y. Sakuraba, K. Hyodo, A. Sakuma, and S. Mitani, Giant anomalous Nernst effect in the $Co_2MnAl_{1-x}$ $Si_x$ Heusler alloy induced by Fermi level tuning and atomic ordering, Phys. Rev. B **101**, 134407 (2020).

[7] T. Yamazaki, T. Seki, R. Modak, K. Nakagawara, T. Hirai, K. Ito, K. Uchida, and K. Takanashi, Thickness dependence of anomalous Hall and Nernst effects in Ni-Fe thin films, Phys. Rev. B **105**, 214416 (2022).

[8] P. Jandl and U. Birkholz, Thermogalvanomagnetic Properties of Sn-Doped $Bi_{95}Sb_5$ and Its Application for Solid State Cooling, J. Appl. Phys. **76**, 7351 (1994).

[9] A. Miura, H. Sepehri-Amin, K. Masuda, H. Tsuchiura, Y. Miura, R. Iguchi, Y. Sakuraba, J. Shiomi, K. Hono, and K. Uchida, Observation of anomalous Ettingshausen effect and large transverse thermoelectric conductivity in permanent magnets, Appl. Phys. Lett. **115**, 222403 (2019).

[10] R. Modak, Y. Sakuraba, T. Hirai, T. Yagi, H. Sepehri-Amin, W. Zhou, H. Masuda, T. Seki, K. Takanashi, T. Ohkubo, and K. Uchida, Sm-Co-based amorphous alloy films for zero-field operation of transverse thermoelectric generation, Sci. Technol. Adv. Mater. **23**, 767 (2022).

[11] S. Noguchi, K. Fujiwara, Y. Yanagi, M. Suzuki, T. Hirai, T. Seki, K. Uchida, and A. Tsukazaki, Bipolarity of Large Anomalous Nernst Effect in Weyl Magnet-Based Alloy Films, Nat. Phys. (2024). https://doi.org/10.1038/s41567-023-02293-z

[12] J. Wang, A. Miura, R. Modak, Y. K. Takahashi, and K. Uchida, Magneto-optical design of anomalous Nernst thermopile, Sci. Rep. **11**, 11228 (2021).

[13] D. Xiao, Y. Yao, Z. Fang, and Q. Niu, Berry-phase effect in anomalous thermoelectric transport, Phys. Rev. Lett. **97**, 026603 (2006).

[14] X. Li, L. Xu, L. Ding, J. Wang, M. Shen, X. Lu, Z. Zhu, and K. Behnia, Anomalous Nernst and Righi-Leduc Effects in $Mn_3Sn$: Berry Curvature and Entropy Flow, Phys. Rev. Lett. **119**, 056601

(2017).

[15] A. Sakai, Y. Mizuta, A. Nugroho, R. Sihombing, T. Koretsune, M. Suzuki, N. Takemori, R. Ishii, D. Nishio-Hamane, R. Arita, P. Goswami, and S. Nakatsuji, Giant anomalous Nernst effect and quantum-critical scaling in a ferromagnetic semimetal, Nat. Phys. **14**, 1119 (2018).

[16] H. Reichlova, R. Schlitz, S. Beckert, P. Swekis, A. Markou, Y. Chen, D. Kriegner, S. Fabretti, G. H. Park, A. Niemann, S. Sudheendra, A. Thomas, K. Nielsch, C. Felser, and S. T. B. Goennenwein1, Large anomalous Nernst effect in thin films of the Weyl semimetal $Co_2MnGa$, Appl. Phys. Lett. **113**, 212405 (2018).

[17] H. Nakayama, K. Masuda, J. Wang, A. Miura, K. Uchida, M. Murata, and Y. Sakuraba, Mechanism of strong enhancement of anomalous Nernst effect in Fe by Ga substitution, Phys. Rev. Mater. **3**, 114412 (2019).

[18] A. Sakai, S. Minami, T. Koretsune, T. Chen, T. Higo, Y. Wang, T. Nomoto, M. Hirayama, S. Miwa, D. Nishio-Hamane, F. Ishii, R. Arita, and S. Nakatsuji, Iron-based binary ferromagnets for transverse thermoelectric conversion, Nature **581**, 53 (2020).

[19] K. Sumida, Y. Sakuraba, K. Masuda, T. Kono, M. Kakoki, K. Goto, W. Zhou, K. Miyamoto, Y. Miura, T. Okuda, and A. Kimura, Spin-polarized Weyl cones and giant anomalous Nernst effect in ferromagnetic Heusler films, Commun. Mater. **1**, 89 (2020).

[20] T. Asaba, V. Ivanov, S. M. Thomas, S. Y. Savrasov, J. D. Thompson, E. D. Bauer, and F. Ronning, Colossal anomalous Nernst effect in a correlated noncentrosymmetric kagome ferromagnet, Sci. Adv. **7**, eabf1467 (2021).

[21] Y. Pan, C. Le, B. He, S. J. Watzman, M. Yao, J. Gooth, J. P. Heremans, Y. Sun, and C. Felser, Giant anomalous Nernst signal in the antiferromagnet $YbMnBi_2$, Nat. Mater. **21**, 203 (2022).

[22] A. Miura, K. Masuda, T. Hirai, R. Iguchi, T. Seki, Y. Miura, H. Tsuchiura, K. Takanashi, and K. Uchida, High-temperature dependence of anomalous Ettingshausen effect in $SmCo_5$-type permanent magnets, Appl. Phys. Lett. **117**, 082408 (2020).

[23] K. Uchida, Anisotropy boosts transverse thermoelectrics, Nat. Mater. **21**, 136 (2022).

[24] K. Uchida, T. Kikkawa, T. Seki, T. Oyake, J. Shiomi, Z. Qiu, K. Takanashi, and E. Saitoh, Enhancement of anomalous Nernst effects in metallic multilayers free from proximity-induced magnetism, Phys. Rev. B **92**, 094414 (2015).

[25] C. Fang, C. H. Wan, Z. H. Yuan, L. Huang, X. Zhang, H. Wu, Q. T. Zhang, and X. F. Han, Scaling relation between anomalous Nernst and Hall effect in $[Pt/Co]_n$ multilayers, Phys. Rev. B **93**, 054420 (2016).

[26] T. Seki, Y. Sakuraba, K. Masuda, A. Miura, M. Tsujikawa, K. Uchida, T. Kubota, Y. Miura, M. Shirai, and K. Takanashi, Enhancement of the anomalous Nernst effect in Ni/Pt superlattices, Phys. Rev. B **103**, L020402 (2021).

[27] R. Kitaura, T. Ishibe, H. Sharma, M. Mizuguchi, and Y. Nakamura, Nanostructure design for high

performance thermoelectric materials based on anomalous Nernst effect using metal/semiconductor multilayer, Appl. Phys. Express **14**, 075002 (2021).

[28] J. Wang, Y. Lau, W. Zhou, T. Seki, Y. Sakuraba, T. Kubota, K. Ito, and K. Takanashi, Strain-Induced Large Anomalous Nernst Effect in Polycrystalline $Co_2MnGa/AlN$ Multilayers, Adv. Electron. Mater. **8**, 2101380 (2022).

[29] Y. Z. Wang, X. M. Luo, Y. Zhang, C. Fang, M. K. Zhao, W. Q. He, G. Q. Yu, C. H. Wan, and X. F. Han, Antiferromagnetic-Metal/Ferromagnetic-Metal Periodic Multilayers for On-Chip Thermoelectric Generation, Phys. Rev. Appl. **17**, 024075 (2022).

[30] K. Uchida, S. Takahashi, K. Harii, J. Ieda, W. Koshibae, K. Ando, S. Maekawa, and E. Saitoh, Observation of the spin Seebeck effect, Nature **455**, 778 (2008).

[31] C. M. Jaworski, J. Yang, S. MacK, D. D. Awschalom, J. P. Heremans, and R. C. Myers, Observation of the spin-Seebeck effect in a ferromagnetic semiconductor, Nat. Mater. **9**, 898 (2010).

[32] K. Uchida, J. Xiao, H. Adachi, J. Ohe, S. Takahashi, J. Ieda, T. Ota, Y. Kajiwara, H. Umezawa, H. Kawai, G. E. W. Bauer, S. Maekawa, and E. Saitoh, Spin Seebeck insulator, Nat. Mater. **9**, 894 (2010).

[33] K. Uchida, H. Adachi, T. Ota, H. Nakayama, S. Maekawa, and E. Saitoh, Observation of longitudinal spin-Seebeck effect in magnetic insulators, Appl. Phys. Lett. **97**, 172505 (2010).

[34] A. Slachter, F. L. Bakker, J. Adam, and B. J. Van Wees, Thermally driven spin injection from a ferromagnet into a non-magnetic metal, Nat. Phys. **6**, 879 (2010).

[35] E. Saitoh, M. Ueda, H. Miyajima, and G. Tatara, Conversion of spin current into charge current at room temperature: Inverse spin-Hall effect, Appl. Phys. Lett. **88**, 182509 (2006).

[36] T. Kikkawa, K. Uchida, S. Daimon, Y. Shiomi, H. Adachi, Z. Qiu, D. Hou, X.-F. Jin, S. Maekawa, and E. Saitoh, Separation of longitudinal spin Seebeck effect from anomalous Nernst effect: Determination of origin of transverse thermoelectric voltage in metal/insulator junctions, Phys. Rev. B **88**, 214403 (2013).

[37] X. Chen, Z. Zhou, Y.-H. Lin, and C. Nan, Thermoelectric thin films: Promising strategies and related mechanism on boosting energy conversion performance, J. Materiomics **6**, 494 (2020).

[38] S. Kommandur and S. Yee, A suspended 3-omega technique to measure the anisotropic thermal conductivity of semiconducting polymers, Rev. Sci. Instrum. **89**, 114905 (2018).

[39] N.-W. Park, W.-Y. Lee, Y.-S. Yoon, G.-S. Kim, Y.-G. Yoon, and S.-K. Lee, Achieving Out-of-Plane Thermoelectric Figure of Merit $ZT$ = 1.44 in a p-Type $Bi_2Te_3/Bi_{0.5}Sb_{1.5}Te_3$ Superlattice Film with Low Interfacial Resistance, ACS Appl. Mater. Interfaces **11**, 38247 (2019).

[40] W.-Y. Lee, M.-S. Kang, N.-W. Park, G.-S. Kim, H.-W. Jang, E. Saitoh, and S.-K. Lee, Phase and Composition Tunable Out-of-Plane Seebeck Coefficients for $MoS_2$-Based Films, ACS Appl. Electron. Mater **4**, 1576 (2022).

[41] D. G. Cahill, Thermal conductivity measurement from 30 to 750 K: The 3ω method, Rev. Sci. Instrum. **61**, 802 (1990).

[42] D. G. Cahill, Analysis of heat flow in layered structures for time-domain thermoreflectance, Rev. Sci. Instrum. **75**, 5119 (2004).

[43] N. Taketoshi, T. Baba, and A. Ono, Electrical delay technique in the picosecond thermoreflectance method for thermophysical property measurements of thin films, Rev. Sci. Instrum. **76**, 094903 (2005).

[44] A. J. Schmidt, R. Cheaito, and M. Chiesa, A frequency-domain thermoreflectance method for the characterization of thermal properties, Rev. Sci. Instrum. **80**, 094901 (2009).

[45] Y. Nakamura, M. Isogawa, T. Ueda, S. Yamasaka, H. Matsui, J. Kikkawa, S. Ikeuchi, T. Oyake, T. Hori, J. Shiomi, and A. Sakai, Anomalous reduction of thermal conductivity in coherent nanocrystal architecture for silicon thermoelectric material, Nano Energy **12**, 845 (2015).

[46] R. B. Wilson and D. G. Cahill, Experimental Validation of the Interfacial Form of the Wiedemann-Franz Law, Phys. Rev. Lett. **108**, 255901 (2012).

[47] J. Kimling, R. B. Wilson, K. Rott, J. Kimling, G. Reiss, and D. G. Cahill, Spin-Dependent Thermal Transport Perpendicular to the Planes of Co/Cu Multilayers, Phys. Rev. B **91**, 144405 (2015).

[48] H. Nakayama, B. Xu, S. Iwamoto, K. Yamamoto, R. Iguchi, A. Miura, T. Hirai, Y. Miura, Y. Sakuraba, J. Shiomi, and K. Uchida, Above-Room-Temperature Giant Thermal Conductivity Switching in Spintronic Multilayers, Appl. Phys. Lett. **118**, 042409 (2021).

[49] A. Sola, M. Kuepferling, V. Basso, M. Pasquale, T. Kikkawa, K. Uchida, and E. Saitoh, Evaluation of thermal gradients in longitudinal spin Seebeck effect measurements, J. Appl. Phys. **117**, 17C510 (2015).

[50] A. Sola, P. Bougiatioti, M. Kuepferling, D. Meier, G. Reiss, M. Pasquale, T. Kuschel, and V. Basso, Longitudinal spin Seebeck coefficient: heat flux vs. temperature difference method, Sci. Rep. **7**, 46752 (2017).

[51] Y. Yamashita, K. Honda, T. Yagi, J. Jia, N. Taketoshi, and Y. Shigesato, Thermal conductivity of hetero-epitaxial ZnO thin Films on *c*- and *r*-plane sapphire substrates: Thickness and grain size effect, J. Appl. Phys. **125**, 035101 (2019).

[52] K. Kobayashi and T. Baba, Extension of the Response Time Method and the Areal Heat Diffusion Time Method for One-Dimensional Heat Diffusion after Impulse Heating: Generalization Considering Heat Sources inside of Multilayer and General Boundary Conditions, Jpn. J. Appl. Phys. **48**, 05EB05 (2009).

[53] See Supplemental Material for RHEED patterns, step model for reproducing XRD pattern of multilayer, and details of TDTR method.

[54] T. Seki, R. Iguchi, K. Takanashi, and K. Uchida, Visualization of anomalous Ettingshausen effect

in a ferromagnetic film: Direct evidence of different symmetry from spin Peltier effect, Appl. Phys. Lett. **112**, 152403 (2018).

[55] T. Seki, R. Iguchi, K. Takanashi, and K. Uchida, Relationship between anomalous Ettingshausen effect and anomalous Nernst effect in an FePt thin film, J. Phys. D: Appl. Phys. **51**, 254001 (2018).

[56] R. Ramos, T. Kikkawa, M. H. Aguirre, I. Lucas, A. Anadón, T. Oyake, K. Uchida, H. Adachi, J. Shiomi, P. A. Algarabel, L. Morellón, S. Maekawa, E. Saitoh, and M. R. Ibarra, Unconventional scaling and significant enhancement of the spin Seebeck effect in multilayers, Phys. Rev. B **92**, 220407 (2015).

TABLE 1 Summarized thickness values obtained by STEM images.

| | | $t_{ML}$ or $t_{alloy}$ (nm) | $t_{Al}$ (nm) |
|---|---|---|---|
| Fe/Pt multilayer | $N = 50$ | 209±3 | 47±6 |
| | $N = 100$ | 223±3 | 50±6 |
| | $N = 200$ | 232±2 | 48±4 |
| FePt alloy film | | 216±2 | 48±3 |

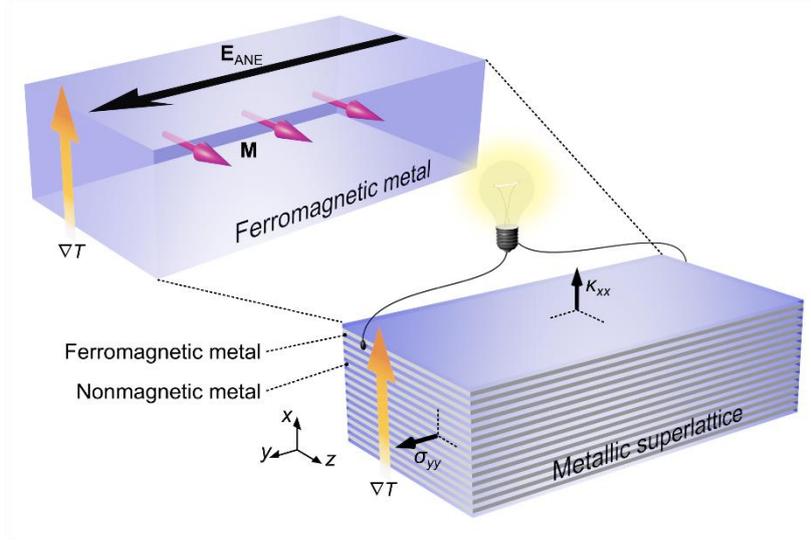

FIG. 1 Schematic illustration of transverse thermoelectric conversion due to the anomalous Nernst effect (ANE) in a metallic multilayer. $\mathbf{E}_{ANE}$, $\mathbf{M}$, $\nabla T$, $\sigma_{yy}$, and $\kappa_{xx}$ denote the ANE-driven electric field, magnetization, temperature gradient, longitudinal electrical conductivity, and thermal conductivity, respectively.

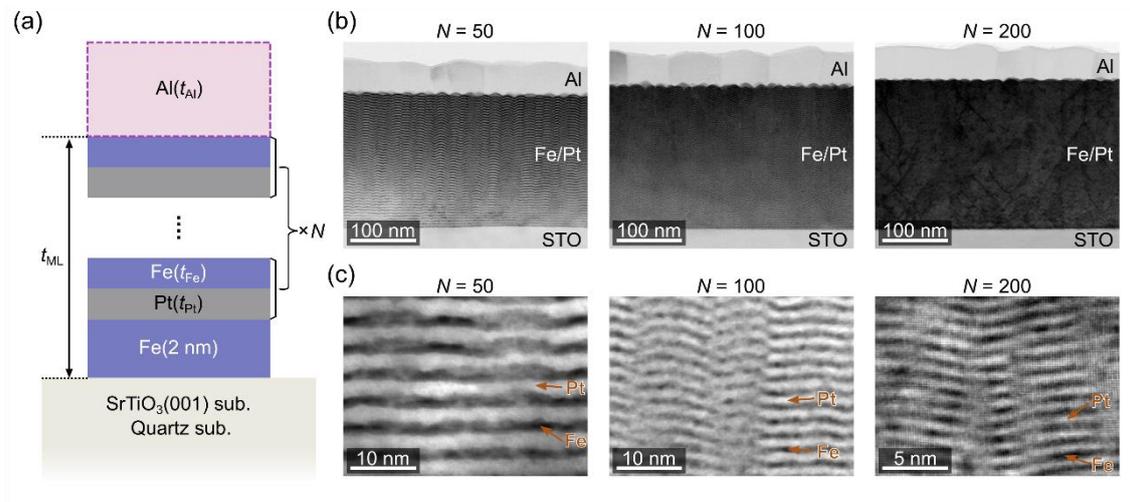

FIG. 2 (a) Layered structure of Fe/Pt metallic multilayers on SrTiO₃ (STO)(001) and quartz substrates. (b) Cross-sectional BF-STEM images of Fe/Pt metallic multilayers on STO substrates for the overall sample cross section. (c) Cross-sectional HAADF-STEM images of Fe/Pt multilayers on STO substrates with higher resolution.

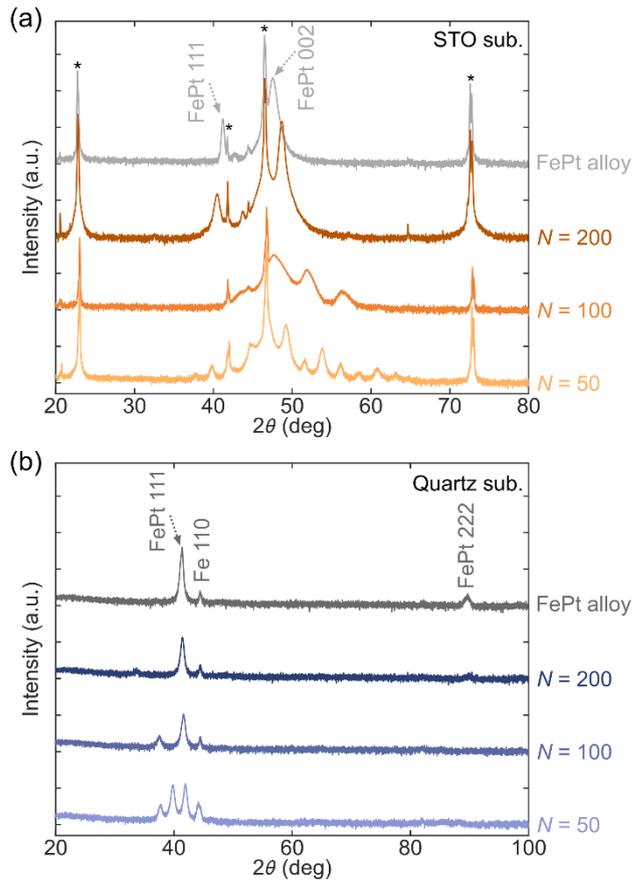

FIG. 3 Out-of-plane X-ray diffraction profiles for Fe/Pt metallic multilayers with $N$ = 50, 100, and 200 and FePt alloy film on (a) STO and (b) quartz substrates. The asterisks in Fig. 3(a) denote the reflections from the STO substrates.

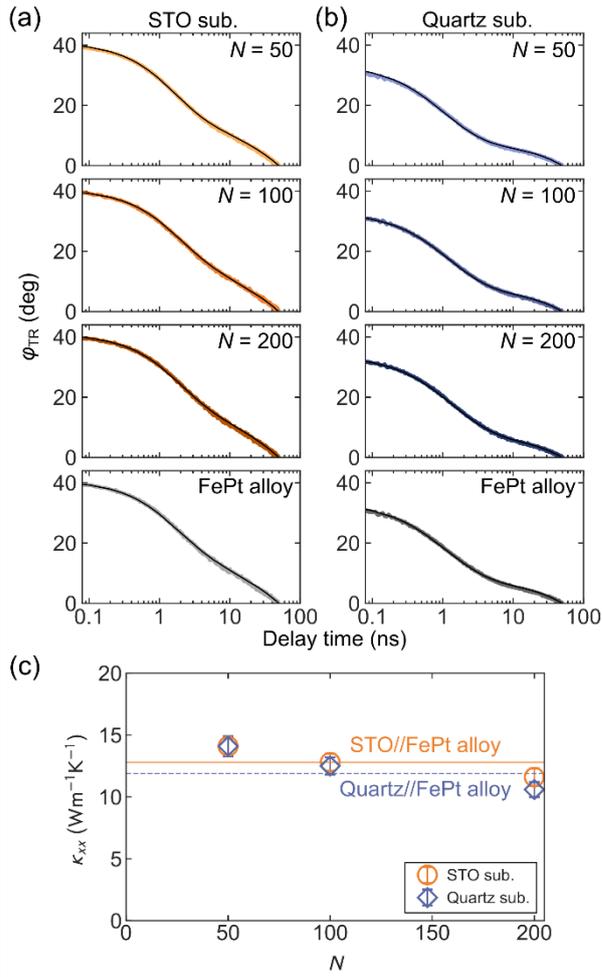

FIG. 4 (a),(b) TDTR phase data $\varphi_{TR}$ vs delay time for Fe/Pt metallic multilayers with $N$ = 50, 100, and 200 and FePt alloy film on (a) STO and (b) quartz substrates. Solid lines are the best fit curves based on one-dimensional heat diffusion model. (c) Number of stacking repetition $N$ dependence of out-of-plane thermal conductivity $\kappa_{xx}$ for Fe/Pt metallic multilayers on STO and quartz substrates. Solid (dashed) line shows the $\kappa_{xx}$ value for FePt alloy film on the STO (quartz) substrate. Error bars represent the uncertainties considering the error propagation.

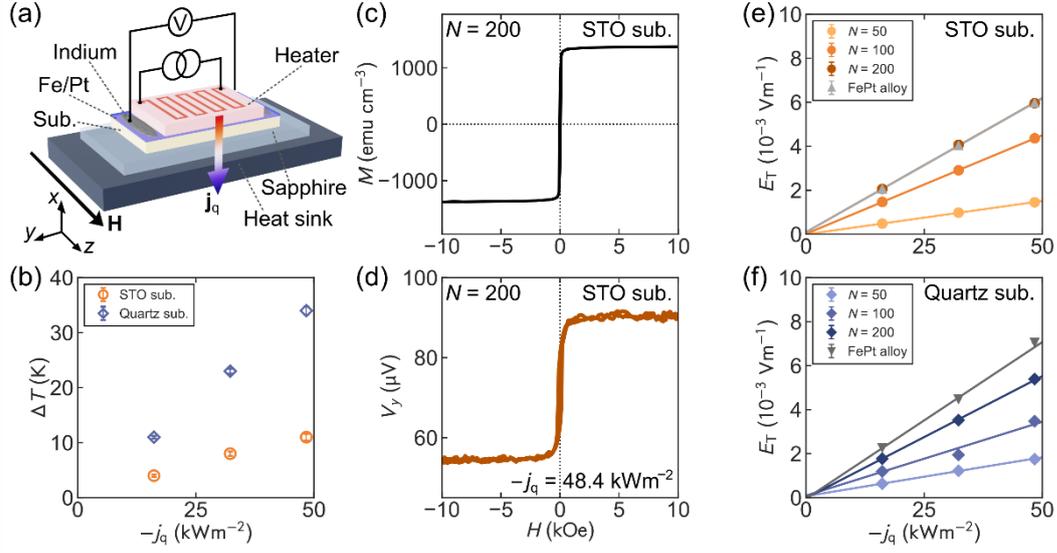

FIG. 5 (a) Schematic illustration of heat flux method. **H** and **j**$_q$ denote the magnetic field and heat flux, respectively. (b) Temperature rise of the heater surface $\Delta T$ as a function of applied heat flux $-j_q$ during the heat flux measurement. Error bars represent the standard deviations of the temperature data at the part of heater surface as captured by the infrared camera. (c), (d) Magnetic field $H$ dependence of (c) magnetization $M$ and (d) voltage along $y$ direction $V_y$ for Fe/Pt multilayer on the STO substrate with $N = 200$. The offset in (d) is due to the parasitic thermopower caused by the Seebeck effect at the electrodes for detecting $V_y$. The present measurement setup has several electrical contacts with different metals. In case that there is a slight temperature difference between the electrodes, the Seebeck effect at the electrodes appears as a $H$-independent thermopower. (e),(f) transverse electric field $E_T$ as a function of $-j_q$ for Fe/Pt multilayers and FePt alloy films on (e) STO and (f) quartz substrates. Solid lines represent the linear fitted lines. Error bars represent the uncertainties considering the error propagation.

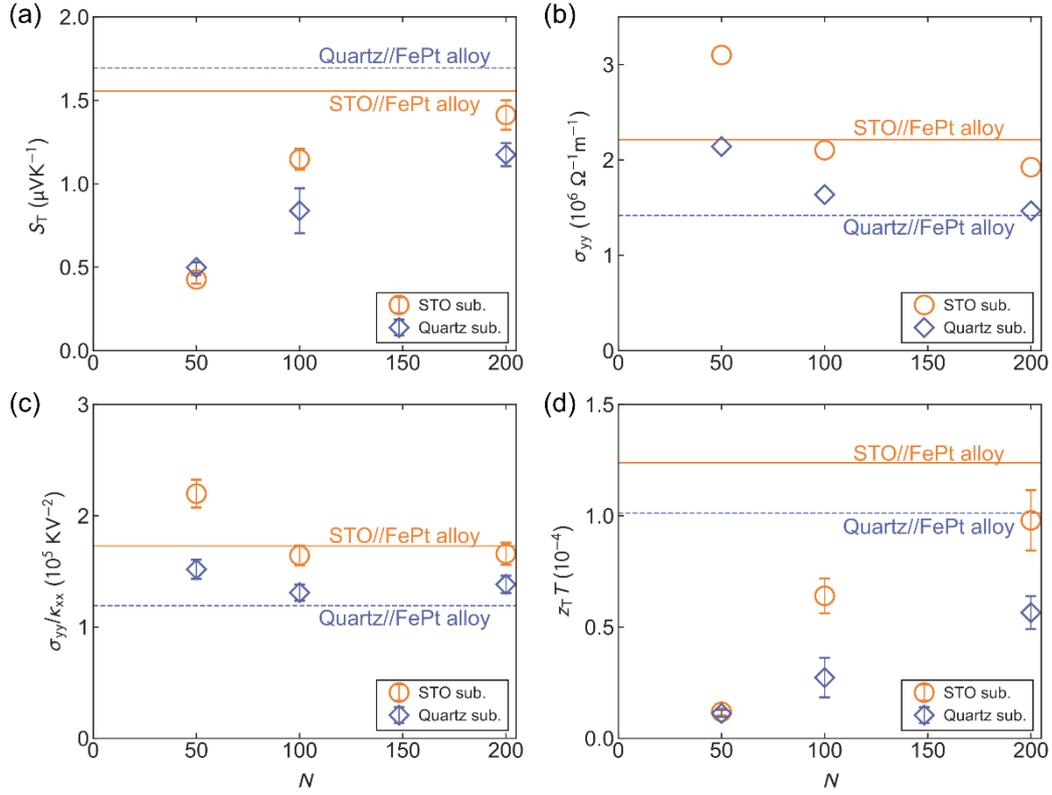

FIG. 6 $N$ dependence of (a) transverse thermoelectric coefficient $S_\mathrm{T}$, (b) electrical conductivity $\sigma_{yy}$, (c) $\sigma_{yy}/\kappa_{xx}$, and (d) figure of merit $z_\mathrm{T}T$ for Fe/Pt metallic multilayers on STO and quartz substrates. Solid (dashed) lines represent the measurement results for FePt alloy films on the STO (quartz) substrate. Error bars in (a), (c), and (d) represent the uncertainty, where the error propagation is considered.

# Supplemental Material for "Quantitative measurement of figure of merit for transverse thermoelectric conversion in Fe/Pt metallic multilayers"

Takumi Yamazaki,[1,*,§] Takamasa Hirai,[2,†,§] Takashi Yagi,[3] Yuichiro Yamashita,[3] Ken-ichi Uchida,[1,2] Takeshi Seki,[1,2,‡] and Koki Takanashi[4,5]

[1] *Institute for Materials Research, Tohoku University, Sendai 980-8577, Japan*

[2] *National Institute for Materials Science, Tsukuba 305-0047, Japan*

[3] *National Institute of Advanced Industrial Science and Technology, Tsukuba 305-8563, Japan*

[4] *Advanced Science Research Center, Japan Atomic Energy Agency, Tokai, Ibaraki 319-1195, Japan*

[5] *Advanced Institute for Materials Research, Tohoku University, Sendai, 980-8577, Japan*

[*] takumi.yamazaki.d5@tohoku.ac.jp

[†] HIRAI.Takamasa@nims.go.jp

[‡] takeshi.seki@tohoku.ac.jp

[§] These authors contributed equally to this work.

## S1. Reflection high-energy electron diffraction (RHEED) patterns for Fe/Pt multilayers

Figure S1 displays the typical reflection high-energy electron diffraction (RHEED) patterns for Fe/Pt multilayer on STO and quartz substrates with $N = 200$. The diffraction patterns were observed just after the growth of the topmost Fe layer. For the Fe/Pt multilayer on the STO substrate, the period of spotty streaks varies with the azimuth relative to the electron beam direction, indicating that the Fe and Pt layers are epitaxially grown in the (001) plane. On the other hand, the multilayer on the quartz substrate exhibits broad spots and ring-shaped backgrounds, suggesting polycrystalline growth of Fe and Pt films.

## S2. Step model for reproducing XRD pattern of Fe/Pt multilayer

The step model is applicable to explain the peak positions for the experimental XRD profiles for the multilayers [1,2]. The X-ray scattering intensity $I(Q)$ for the Fe/Pt multilayer is given by

$$I(Q) = I_e |F_{\text{Fe}}(Q) + F_{\text{Pt}}(Q)\exp(iQt_{\text{Fe}})|^2 \left| \sum_{k=0}^{N-1} \exp(iQk\Lambda) \right|^2, \qquad \text{(S1)}$$

where $I_e$ is Thomson scattering intensity, $F_{\text{Fe(Pt)}}(Q)$ is the structural factor of Fe (Pt) layer, $\Lambda$ is the multilayer period, and $Q$ is the scattering vector. $\left| \sum_{k=0}^{N-1} \exp(iQk\Lambda) \right|^2$ corresponds to Laue function $L(Q)$, and is expressed as

$$L(Q) = \left| \sum_{k=0}^{N-1} \exp(iQk\Lambda) \right|^2 = \frac{\sin^2\left(\frac{NQ\Lambda}{2}\right)}{\sin^2\left(\frac{Q\Lambda}{2}\right)}, \tag{S2}$$

where $N$ is the number of stacking repetition. The term of structural factor on the right-hand side of Eq. (S1) is expressed as

$$|F_{\text{Fe}}(Q) + F_{\text{Pt}}(Q)\exp(iQt_{\text{Fe}})|^2 =$$
$$|F_{\text{Fe}}(Q)|^2 + |F_{\text{Pt}}(Q)|^2 + F_{\text{Fe}}(Q)F_{\text{Pt}}{}^*(Q)\exp(-iQt_{\text{Fe}}) + F_{\text{Fe}}{}^*(Q)F_{\text{Pt}}(Q)\exp(iQt_{\text{Fe}}). \tag{S3}$$

Using the atomic scattering factor ($f_{\text{Fe}}$ and $f_{\text{Pt}}$), areal atomic density ($\eta_{\text{Fe}}$ and $\eta_{\text{Pt}}$), lattice spacing ($d_{\text{Fe}}$ and $d_{\text{Pt}}$), and number of lattice plane ($n_{\text{Fe}}$ and $n_{\text{Pt}}$), the term of structural factor can be transformed into

$$|F_{\text{Fe}}(Q) + F_{\text{Pt}}(Q)\exp(iQt_{\text{Fe}})|^2 = f_{\text{Fe}}{}^2(Q)\eta_{\text{Fe}}{}^2 \frac{\sin^2\left(\frac{n_{\text{Fe}}Qd_{\text{Fe}}}{2}\right)}{\sin^2\left(\frac{Qd_{\text{Fe}}}{2}\right)} + f_{\text{Pt}}{}^2(Q)\eta_{\text{Pt}}{}^2 \frac{\sin^2\left(\frac{n_{\text{Pt}}Qd_{\text{Pt}}}{2}\right)}{\sin^2\left(\frac{Qd_{\text{Pt}}}{2}\right)}$$

$$+2f_{\text{Fe}}(Q)f_{\text{Pt}}(Q)\eta_{\text{Fe}}\eta_{\text{Pt}} \frac{\sin\left(\frac{n_{\text{Fe}}Qd_{\text{Fe}}}{2}\right)}{\sin\left(\frac{Qd_{\text{Fe}}}{2}\right)} \frac{\sin\left(\frac{n_{\text{Pt}}Qd_{\text{Pt}}}{2}\right)}{\sin\left(\frac{Qd_{\text{Pt}}}{2}\right)}\cos\left(\frac{\Lambda Q}{2}\right). \tag{S4}$$

Figure S2 shows the comparison between the experimental XRD profile for $N = 50$ on STO and the calculated XRD profile. The experimental peak positions are reproduced by the calculation, indicating the well-defined structure of the present Fe/Pt multilayer.

### S3. Details of time-domain thermoreflectance (TDTR) method

Our time-domain thermoreflectance (TDTR) measurement system with front-heating and front-detection configuration consists of two lasers, and the delay time between the pump and probe lasers is electrically controlled [3,4]. The repetition rate, pulse duration, central wavelength, and $1/e^2$ spot diameter of pump (probe) laser were 20 MHz (20 MHz), 0.5 ps (0.5 ps), 1550 nm (775 nm), and ~90 μm (~30 μm), respectively. The intensity of pump laser is modulated with a frequency of 200 kHz for lock-in detection.

Following to conventional TDTR experiments [5], some parameters are fixed besides the unknown thermal transport parameter of sample layer. In this study, the following parameters are the fixed values: volumetric heat capacity $\rho C$ ($2.44\times10^6$ J m$^{-3}$ K$^{-1}$ [6]) and thermal diffusivity $D$ ($9.8\times10^{-5}$ m$^2$ s$^{-1}$ [7]) of Al, $\rho C$ of Fe/Pt multilayer and FePt alloy film ($3.13\times10^6$ J m$^{-3}$ K$^{-1}$), thermal effusivity of STO (4800 J m$^{-2}$ s$^{-0.5}$ K$^{-1}$) and quartz (1450 J m$^{-2}$ s$^{-0.5}$ K$^{-1}$), and thickness of Al, Fe/Pt multilayer, and FePt alloy film (see Table 1 in the main text). $\rho C$ of Fe/Pt multilayer and FePt alloy film was estimated by calculating the weighted average of the bulk $\rho C$ values of Fe and Pt [6], where $\rho$ is density and $C$ is specific heat. The thermal effusivity (= $\rho C D^{0.5}$) of STO and quartz substrates was verified by individually measuring $D$ using the laser flash method, $C$ using the differential scanning calorimetry, and $\rho$ using the Archimedes method by using plain STO and quartz substrates. The thickness of each layer was estimated from the STEM images. The out-of-plane thermal diffusivity $D_{xx}$ of Fe/Pt multilayer and FePt alloy film, interfacial thermal resistance $R_1$ between the Al layer and Fe/Pt multilayer or FePt alloy film (summarized in Fig. S3 as a function of $N$), and interfacial thermal resistance $R_2$ (= $4.0 \times 10^{-9}$ m$^2$ K W$^{-1}$) between Fe/Pt multilayer or FePt alloy film and substrates, are the fitting variables. From the obtained $D_{xx}$ and given $\rho C$ of Fe/Pt multilayer and FePt alloy film, out-of-plane $\kappa_{xx}$ (=$\rho C D_{xx}$) of Fe/Pt multilayer and FePt alloy film can be evaluated.

To ensure the validity of our TDTR method, we performed $\kappa_{xx}$ measurements using pure-Fe films. Figure S4 shows the measured TDTR data and fitting curves for the 200-nm and 250-nm-thick Fe films on STO substrates. Faster decay of temporal response than that in Fe/Pt multilayers [see Fig. 5(a) in the main text] reflects much larger $\kappa_{xx}$ of the Fe film. The analysis of TDTR signals reveals $\kappa_{xx}$ = 81 W m$^{-1}$ K$^{-1}$ for both Fe films, which is similar to the thermal conductivity of bulk Fe (80.4 W m$^{-1}$ K$^{-1}$ [8]), confirming the reliability of our $\kappa_{xx}$ measurement using TDTR technique.

Multiple measurements with different laser irradiation positions were performed to confirm the reproducibility of the TDTR method. Figure S5 depicts the TDTR phase data $\varphi_{TR}$ vs delay time for the Fe/Pt multilayer with $N = 200$ on the STO substrate at three different laser irradiation positions. As shown in Fig. S5, experimental curves with the same shape were obtained, indicating the high reproducibility of the TDTR method.

**References**

[1] A. Segmüller and A. E. Blakeslee, X-ray Diffraction from One-Dimensional Superlattices in GaAs$_{1-x}$P$_x$ Crystals, J. Appl. Crystallogr. **6**, 19 (1973).

[2] I. K. Schuller, New Class of Layered Materials, Phys. Rev. Lett. **44**, 1597 (1980).

[3] Y. Yamashita, K. Honda, T. Yagi, J. Jia, N. Taketoshi, and Y. Shigesato, Thermal conductivity of hetero-epitaxial ZnO thin Films on $c$- and $r$-plane sapphire substrates: Thickness and grain size effect, J. Appl. Phys. **125**, 035101 (2019).

[4] R. Modak, Y. Sakuraba, T. Hirai, T. Yagi, H. Sepehri-Amin, W. Zhou, H. Masuda, T. Seki, K. Takanashi, T. Ohkubo, and K. Uchida, Sm-Co-based amorphous alloy films for zero-field operation of transverse thermoelectric generation, Sci. Technol. Adv. Mater. **23**, 767 (2022).

[5] P. Jiang, X. Qian, and R. Yang, Tutorial: Tutorial: Time-domain thermoreflectance (TDTR) for thermal property characterization of bulk and thin film materials, J. Appl. Phys. **124**, 161103 (2018).

[6] W. M. Haynes, CRC Handbook of Chemistry and Physics, 95th ed. (CRC Press, Boca Raton, FL, 2014-2015).

[7]  Thermal diffusivity $D$ of Al is derived using the relation of $D=\kappa/\rho C$, where $\kappa$ (240 W m$^{-1}$ K$^{-1}$ [4]) is thermal conductivity.

[8]  G. K. White and M. L. Minges, Thermophysical properties of some key solids: An update, Int. J. Therm. **18**, 1269 (1997).

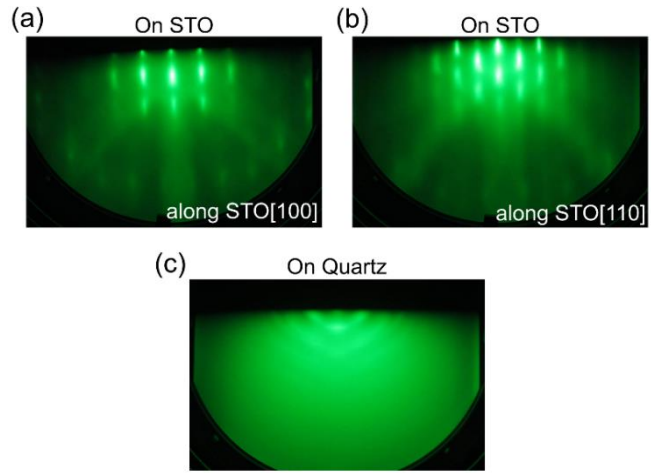

FIG. S1 RHEED patterns for Fe/Pt multilayers with $N = 200$ (a) on the STO substrate with azimuth of STO[100], (b) on the STO substrate with azimuth of STO[110], and (c) on the quartz substrate.

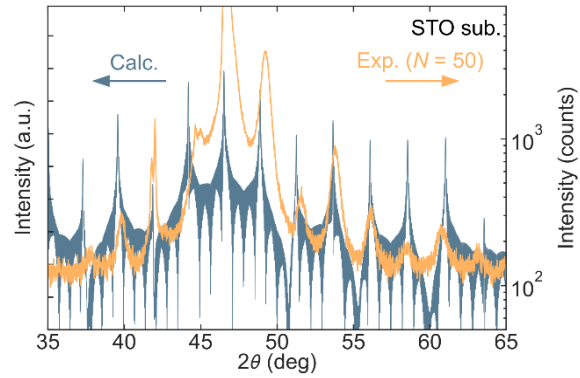

FIG. S2 Experimental XRD profile for $N = 50$ on STO and calculated XRD profile based on the Laue function with the assumption of the step model for the superstructure.

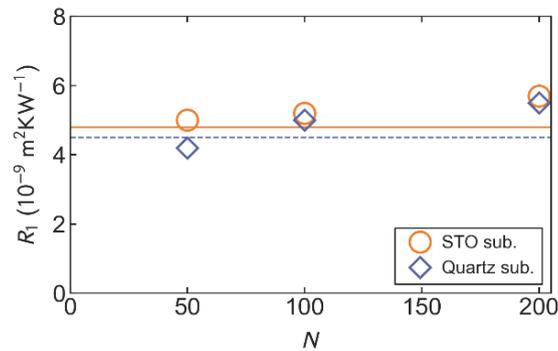

FIG. S3 $N$ dependence of $R_1$ for Fe/Pt metallic multilayers on STO and quartz substrates. Solid (dashed) line shows the $R_1$ value for FePt alloy film on the STO (quartz) substrate.

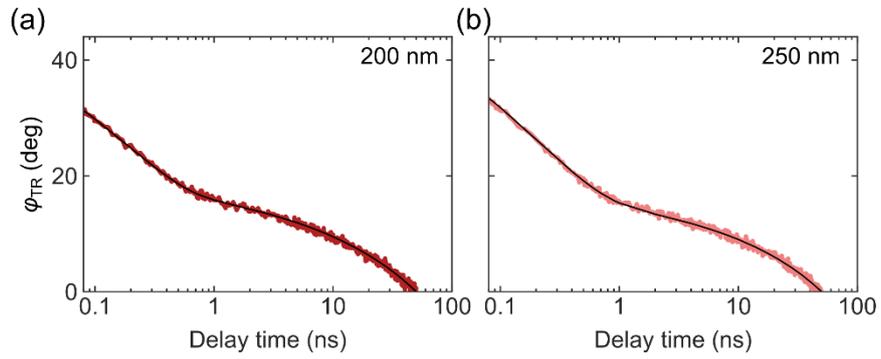

FIG. S4 TDTR phase data $\varphi_{TR}$ vs delay time for (a) 200-nm and (b) 250-nm-thick Fe films on STO substrates. Solid lines are the best fit curves based on one-dimensional heat diffusion model.

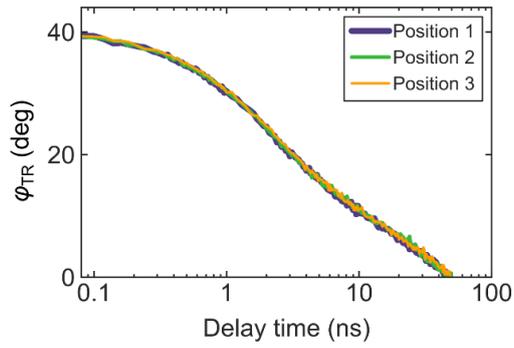

FIG. S5 $\varphi_{TR}$ vs delay time for the Fe/Pt multilayer with $N = 200$ on the STO substrate at different laser irradiation positions.